\documentclass[galaxies,article,accept,oneauthor,pdftex,10pt,a4paper]{mdpi} 

\usepackage{color}

\firstpage{1} 
\articlenumber{x}
\doinum{10.3390/------}
\pubvolume{xx}
\pubyear{2016}
\copyrightyear{2016}
\externaleditor{Academic Editor: name}
\history{Received: date; Accepted: date; Published: date}
\pdfoutput=1

 \theoremstyle{mdpi}
 \newcounter{thm}
 \newcounter{ex}
 \newcounter{re}
 \theoremstyle{mdpidefinition}

\Title{Gaussian Processes for Blazar Variability Studies}

\Author{Vassilis Karamanavis}
\AuthorNames{Vassilis Karamanavis}
\address{%
\quad Max-Planck-Institut f\"ur Radioastronomie, Auf dem H\"ugel 69, 
D-53121 Bonn, Germany; vkaramanavis@mpifr.de
}

\abstract{This article shortly introduces Gaussian processes as a new approach 
for modelling time series in the field of blazar physics. In the second part of 
the paper, recent results from an application of GP modelling to the 
multi-wavelength light curves of the blazar PKS\,1502+106 are discussed.}


\begin{document}

%%%%%%%%%%%%%%%%%%%%%%%%%%%%%%%%%%%%%%%%%% 
%% Only for the journal Gels: Please place the  Experimental Section after the 
% Conclusions

%%%%%%%%%%%%%%%%%%%%%%%%%%%%%%%%%%%%%%%%%%
% \setcounter{section}{-1} %% Remove this when starting to work on the template.

\section{Introduction}

Detailed investigations of the emission from blazars have offered, and 
continue to offer, important insights into this extreme population of 
active galactic nuclei (AGN). 
% With their relativistic jets almost aligned with 
% our line of sight, blazars comprise an AGN population with properties such as 
Blazar variability is an indispensable tool because it offers information on 
scales even smaller than those probed with imaging techniques. 
Furthermore, given the predictions of various blazar variability models 
\citep[e.g.,][]{1985ApJ...298..114M, 
1992A&A...254...71V,2001MNRAS.325.1559S,1992A&A...255...59C}
% ,2001Sci...294..128L}
, we are able to assess their validity taking advantage of the 
fingerprints of 
each model in the time and frequency domains, namely on light curves and 
broadband spectra.

Numerous monitoring efforts have been in place since the early days of the 
discovery of AGN and blazars, however most, if not all, 
lacked the high time resolution required for detailed studies. The tables have 
turned with the launch of the \textit{Fermi} $\gamma$-ray space 
telescope and complementary efforts for multi-frequency 
blazar monitoring. One such example in the radio domain is the F-GAMMA 
($Fermi$-GST AGN Multi-frequency Monitoring Alliance) program
\citep{2007AIPC..921..249F,2010arXiv1006.5610A}. 
The F-GAMMA operated 
between 2007 and 2015 and served as a prime provider of single-dish 
radio monitoring complementary to \textit{Fermi} \cite{2016arXiv160802580F}.

The goal of variability studies is the extraction of parameters such as the 
variability amplitudes, time scales, and others, from observations. 
The problem of modelling the observed variability in the radio and other 
bands, has been addressed using numerous approaches. For modelling light 
curves and the flares seen in them, Gaussian, exponential, and various other 
fitting methods have been used 
(e.g. 
\cite{1999ApJS..120...95V,2011MNRAS.415.1631K,2015A&A...575A..55A}; this is 
by no means a complete list).
Depending on each specific case, 
they work better or worse but the 
common characteristic among all the above is that a functional form of the 
underlying process is always assumed. In other words, a parametric approach is 
being employed. However, this might not be appropriate and we might have to 
admit that at times we have weak prior knowledge as to what is the 
mathematical form of such a process.

An elegant way to overcome this state of limited knowledge is by 
means of Bayesian  inference. Specifically, this paper is concerned with the 
application of  Gaussian processes in the field of blazar variability studies 
and modelling of  time series. In the next section, a brief introduction to 
Gaussian processes is given. Then, an application to the light curves of the 
blazar PKS\,1502+106 is discussed,
% . Namely, the application of GP regression to 
% the radio light curves of the source 
aiming to quantify an outburst seen from 2008 to 2010 and extract 
light curve 
parameters like the flare amplitude and putative delays between its peak at 
different bands.
First results along with an extended discussion can be found in 
\cite{2015PhDT.......232K}.

%%%%%%%%%%%%%%%%%%%%%%%%%%%%%%%%%%%%%%%%%%
\section{Non-parametric models and Gaussian processes}

The problem at hand is as follows: we want to extract (or learn, in 
the context of Bayesian machine learning) the properties of a system from 
a given set of observations. This can be a function $f()$ that describes the 
system well enough and predicts its future evolution. There are essentially two 
ways of achieving this goal. The first involves the selection of a 
class of functions, e.g. linear, quadratic, Gaussian etc. with (free) 
parameters, the 
values of which are obtained through minimizing the residuals between $f()$ and 
our observations. Such a selection can result from strong prior knowledge 
regarding the system. One can increase the flexibility of the model by 
increasing the number of parameters at the expense, though, of potential 
overfitting; i.e. describing random small scale variations of the data 
instead of the general trend.
% Here, the amount of observations only affects the estimation of 
% these  parameters that in turn affect   

Still, cases in which we have little or no prior knowledge regarding the 
underlying process, and consequently the appropriate model to use, exist in 
abundance. Non-parametric models offer an interesting alternative to the first 
approach. By comparison, non-parametric models ``translate'' our prior 
knowledge about a system, no matter how generic this might be, e.g. smooth 
and/or continuous changes, into probability distributions of functions that 
are more likely to describe the system. The task of modelling then reduces to 
fine-tuning these prior distributions so that they include functions that best 
describe our observations; i.e. the posterior distributions. Since no explicit 
use of free parameters is made, these models are referred to as non-parametric. 
Non-parametric models, however, use so-called hyperparameters which instead of  
determining the properties of functions, as in the first approach, 
govern the properties of the distribution. These still need to be inferred and 
Bayesian model selection provides a robust framework for doing so.

Gaussian processes (GPs) are the generalization of uni- or multivariate 
Gaussian distributions of variables into the space of functions. GPs offer a 
very flexible framework for modelling unknown functions 
\cite{Williams96gaussianprocesses, Roberts20110550, citeulike:13629733}. They 
are widely used in machine learning applications and a new field of applications 
in astronomy has recently opened 
\citep[e.g.][]{2012MNRAS.419.2683G, 2012MNRAS.419.3147A, 2015MNRAS.452.2269R, 
2016A&A...590A..48K}. The interested reader is referred to 
\cite{Rasmussen:2005:GPM:1162254} for additional details.

\subsection{Gaussian processes for regression}

\begin{figure}
\centering
\includegraphics[width=\textwidth]{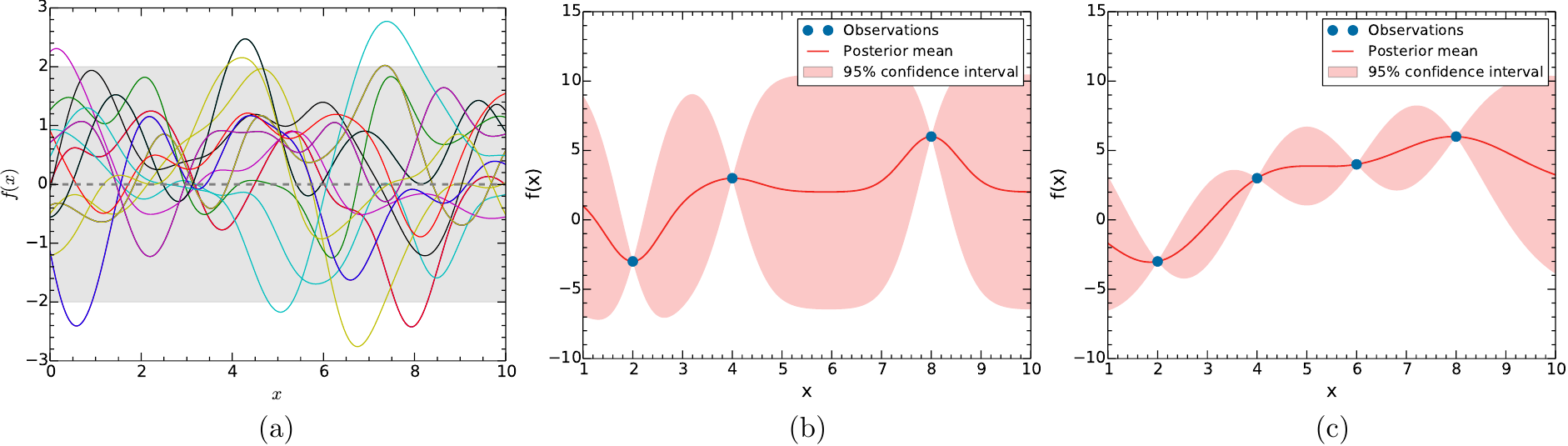}
\caption{Gaussian process prior and posterior distributions: (\textbf{a}) 
Twenty randomly selected samples from the prior distribution, with zero mean 
and $\sigma = 1$, using a squared exponential covariance kernel. 
(\textbf{b}) Regression results after training the GP with three error-free 
data points.
(\textbf{c}) Same as in panel (b), however note the improved result with 
the addition of one more data point.}
\label{fig:1}
\end{figure}

Given a set of observations, e.g. a typical light curve, with the form $D = 
% \{(J_1, t_1), (J_2, t_2), (J_3, t_3)\}$
(x_i, y_i)$, where $y_i$ represent  recorded flux densities at times $x_{i}$, 
we would like to find all functions drawn from the prior distribution 
which 
pass through all observations. The latter distribution would be the posterior 
distribution and its mean value describes the data best (see Figure 
\ref{fig:1}). 

What generates the prior distribution is a covariance kernel or function. This 
governs the similarity (covariance) between any two function values, at 
$x_i$ and $x_j$ say. It is chosen based on our prior knowledge of the system. 
There exists a multitude of kernels which can model unknown functions based 
on their characteristics, e.g. periodic, linear etc. 
\citep[e.g.][]{duvenaud-thesis-2014}. It is noteworthy that sums and 
products of kernels produce valid kernels too. For the purposes of this paper, 
the squared-exponential (SE) kernel is chosen.
This is the simplest kernel and the only underlying assumption is that of a 
smooth process (i.e. the functions it produces have derivatives of all orders 
everywhere in their domain). The SE kernel is defined as
\begin{equation}
 k(x_{i}, x_{j}) = \sigma_{f}^{2} \exp \left[  
\dfrac{-(x_{i}-x_{j})^{2}}{2l^{2}} 
\right],
\label{eq:SqExpKernel}
\end{equation}
with $l$ the the characteristic length scale, i.e. the distance in the $x$ 
dimension after which the function changes significantly and $\sigma_{f}^{2}$, 
the variance, i.e. the mean distance of the function from its mean (serves only 
as a scaling factor). Here, $l$ and $\sigma_{f}^{2}$ are the hyperparameters.

Uncertainty is accommodated by simply adding a term describing it, to the 
noiseless SE covariance kernel of Eq. \ref{eq:SqExpKernel}, which then becomes
\begin{equation}
\boldsymbol{k}_{n} (x_{i}, x_{j}) = \boldsymbol{k} (x_{i}, x_{j}) + \sigma^{2}
\boldsymbol{I},
\end{equation}
with $\boldsymbol{I}$ the identity matrix \citep[e.g.][]{Roberts20110550}.

\subsection{Training the Gaussian process}
Bayesian model selection provides the way for optimizing the hyperparameters, 
thereby refining the posterior distribution and enabling us to perform 
inference. We essentially update our prior knowledge in the light of a data 
set. This is done by means of maximizing the log marginal likelihood 
\citep{Williams96gaussianprocesses}, as outlined below.

For the SE kernel, the vector of hyperparameters is
$\boldsymbol{\theta} = \{ l, \sigma^{2} \}$ and
the probability (or evidence) of the training data $\boldsymbol{y}$,
given the hyperparameters vector $\boldsymbol{\theta}$, is 
% \begin{equation}
 $ p(\boldsymbol{y} \,|\, \boldsymbol{x}, \boldsymbol{\theta} )$.
% \end{equation}
The log marginal likelihood is given by
\begin{equation}
\mathcal{L} = \log  p(\boldsymbol{y} \,|\, \boldsymbol{x}, \boldsymbol{\theta} )
=
 % term 1
 - \dfrac{1}{2} \boldsymbol{y}^{T} \boldsymbol{k}_{n}^{-1} \boldsymbol{y}
 % term 2
 - \dfrac{1}{2} \log |\boldsymbol{k}_{n}|
 % term 3
 - \dfrac{n}{2} \log 2 \pi.
\end{equation}
More generally, in the case of a hyperparameter vector 
$\boldsymbol{\theta} = \{ 
\theta_{j} \,|\, j = 1, ..., n \} $, 
the derivatives of the log marginal likelihood with respect to each 
$\theta_{j}$ are
\begin{equation}
 \dfrac{\partial \mathcal{L}}{\partial \theta_{j}} = 
\dfrac{1}{2} \boldsymbol{y}^{T} \dfrac{\partial
\boldsymbol{k}_{n}}{\partial \theta_{j}} \boldsymbol{k}_{n}^{-1}
\boldsymbol{y}
- \dfrac{1}{2} ~\mbox{Trace} \left( \boldsymbol{k}_{n}^{-1} 
\dfrac{\partial \boldsymbol{k}_{n}}{\partial \theta_{j}} \right).
\label{eq:gradient}
\end{equation}
By means of numerical gradient optimization algorithms one can maximize the log 
marginal likelihood using Eq. \ref{eq:gradient} and obtain the best set of
hyperparameters (for an outline of the algorithm 
see \cite{Rasmussen:2005:GPM:1162254}).

%%%%%%%%%%%%%%%%%%%%%%%%%%%%%%%%%%%%%%%%%%
\section{Application to the light curves of the blazar PKS\,1502+106}

PKS\,1502+106 (S3\,1502+10, OR\,103) is a powerful flat-spectrum radio quasar 
(FSRQ), driven by a supermassive black hole 
% with a mass 
of about $10^{9}$ 
M$_{\odot}$, at redshift $z = 1.8385$ (see \cite{2016A&A...586A..60K} and 
references therein). The source has been detected by \textit{Fermi} in 2008 
showing an intense $\gamma$-ray outburst at MeV--GeV energies 
\citep{2010ApJ...710..810A}. The flare was 
monitored by F-GAMMA and other programs in the radio band.
Radio data from the F-GAMMA 
program\footnote{\url{www.mpifr-bonn.mpg.de/div/vlbi/fgamma/fgamma.html}} were 
employed, including observations with the Effelsberg 100-m (EB) and the IRAM 
30-m (PV) telescopes in ten frequency bands from 2.64 to 142.33\,GHz.
Monthly observations from the EB and PV telescopes were performed 
quasi-simultaneously, thus ensuring maximum spectral coherency. A detailed 
description of the observational setup and data reduction is provided elsewhere 
\citep{2014MNRAS.441.1899F,2015A&A...575A..55A, 
2015Nestoras,2016arXiv160802580F}.
% We also employ data from the Owens Valley Radio Observatory (OVRO) 40-m 
% blazar 
% monitoring program\footnote{\url{www.astro.caltech.edu/ovroblazars}} at 15\,GHz
% \citep{2011ApJS..194...29R} and at 226.5\,GHz from the Submillimeter Array (SMA)
% observer center database \citep{2007ASPC..375..234G}.

Considering the full length of available multi-frequency light curves, GP 
regression was applied to the data at each frequency separately. A variant of 
the software by \cite{scikit-learn} was used, which was adapted to our specific 
modelling needs. A suite of machine learning applications for Python, including 
GP regression, is available online\footnote{\url{www.scikit-learn.org}}.
To obtain the best unbiased result, the length scale 
parameter $l$ was randomly initialized 100 times. Then, the 
posterior mean is returned with a 95\% confidence interval for the flux 
density, along with the set of hyperparameters which maximize the log 
marginal likelihood. The relevant light curve parameters extracted are 
the peak amplitude ($S_{\rm m}$), the time of the peak, 
% $t_{\rm m}$, 
and the cross-band delay.
% $\tau_{\rm GP}$. 
Furthermore, 
the flare rise and decay times 
% ($t_{\rm r}$ and $t_{\rm d}$, respectively) 
are calculated between the peak and the two flux density minima which 
immediately precede and follow 
$S_{\rm m}$ \cite{2016A&A...590A..48K}.
% In Fig. \ref{fig:1} the results of regression at each frequency are visualized.
% The values of $S_{\rm m}$, $t_{\rm m}$, $t_{\rm r}$, and $t_{\rm d}$
% characterizing the flare are visible in Table \ref{tab:1}.
\begin{figure}
\centering
\includegraphics[width=\textwidth]{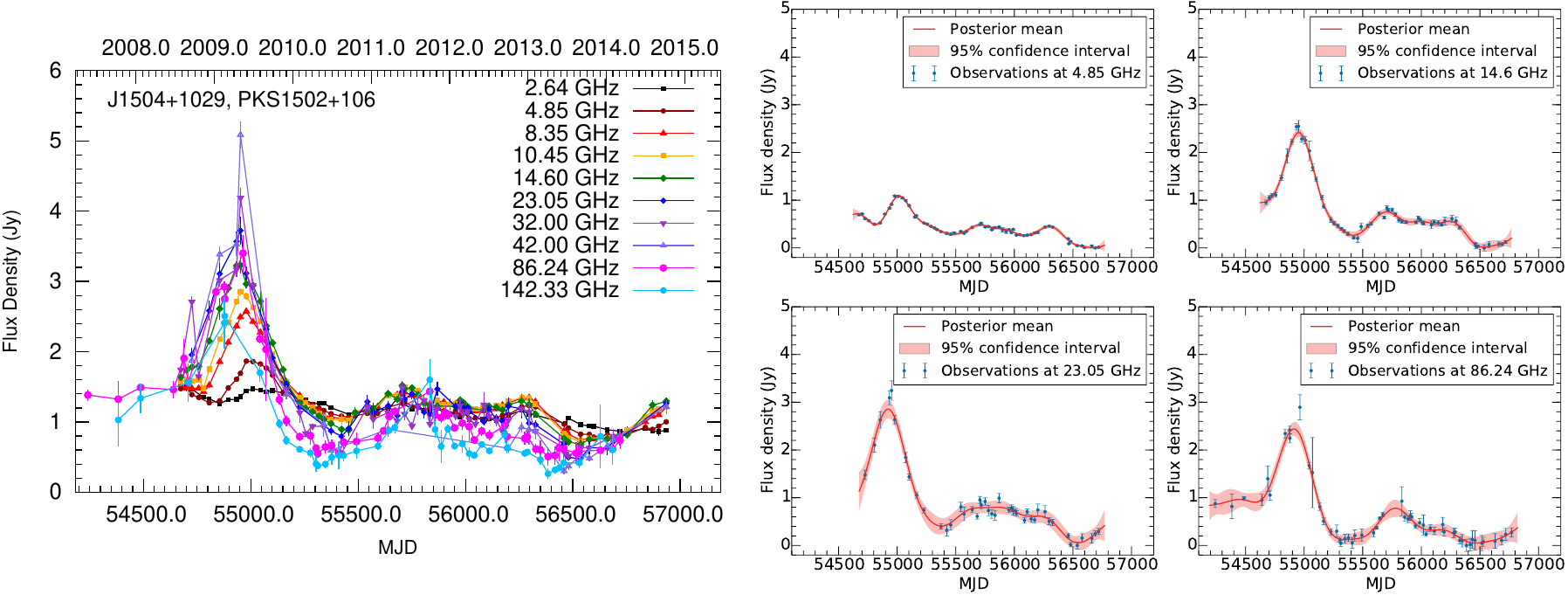}
\caption{Multi-wavelength light curves of PKS\,1502+106. (\textbf{Left}) 
F-GAMMA light curves between 2.64 and 142.33\,GHz. (\textbf{Right}) GP 
regression curves for the data at 4.85, 14.6, 23.05, and 86.24\,GHz. For the 
results at all frequencies 
see \cite{2016A&A...590A..48K}.}
\label{fig:2}
\end{figure}

%%%%%%%%%%%%%%%%%%%%%%%%%%%%%%%%%%%%%%%%%%
\section{Results and discussion}
\subsection{Quantifying the broadband outburst of PKS\,1502+106}
PKS\,1502+106 has shown an isolated outburst visible across the electromagnetic 
spectrum. The results of the GP regression are visible in the right
panels of 
Fig. \ref{fig:2}. The amplitudes at each of our radio frequencies 
first increase, up to about 43\,GHz where they peak, and then drop at higher 
$\nu$ (Fig. 4 of Ref. \cite{2016A&A...590A..48K}).The flare peaks with 
increasing time delay as we follow it towards lower frequencies (Fig. 5 of 
Ref. \cite{2016A&A...590A..48K}). The dependence of amplitudes and 
delays on $\nu$ can be well approximated by power laws, a telltale sign of 
shock evolution within the source's parsec-scale jet. These results are in 
agreement with the study of the source using  mm-wave very-long-baseline 
interferometry (mm-VLBI; see \cite{2016A&A...586A..60K}).

The observed time lags between radio frequencies can be attributed to 
synchrotron opacity effects. In this context, the flare maxima appear 
first at higher frequencies and gradually progress to lower frequencies 
\citep[e.g.][]{1992A&A...254...71V}. Timing the flare maximum at each frequency 
with respect to the data at 142.33\,GHz yields the following delays. The 
longest delay of $(205 \pm 2.0)$\,days is seen between $142.33$ and $2.64$\,GHz. 
Decreasing lags are clearly visible towards higher frequencies with a time 
lag of $(136.9 \pm 8.1)$\,d at 4.85\,GHz and 
$(96.2 \pm 3.8)$\,d at 8.35\,GHz. At higher frequencies, the lags are 
$(85.8 \pm 5.2)$\,d at 10.45\,GHz,
$(67.2 \pm 0.9)$\,d at 14.6,
$(39.4 \pm 5.6)$\,d at 23.03, and
$(39.8 \pm 11.2)$\,d at 32\,GHz.
At the two highest frequencies of 43.05\,GHz and 86.24\,GHz, the lags are
$(40.8 \pm 3.1)$\,d and $(24.8 \pm 0.2)$\,d, respectively (Table 3 of Ref. 
\cite{2016A&A...590A..48K}).

The aforementioned lags allow the calculation of the synchrotron opacity 
profile of PKS\,1502+106. Considering that at a given frequency the 
core represents a feature whose optical depth is close to unity, the 
standard relativistic jet paradigm predicts that the apparent position of this 
unit-opacity surface depends on observing frequency 
(see \cite{2016A&A...590A..48K} and references therein).
The importance of this ``core-shift'' effect lies in the fact that it gives 
critical insights into the physical processes of ultracompact
jets, like the de-projected distance between the core (at each 
frequency) and the jet base, as well as the strength of the magnetic field along 
the flow \citep[e.g.,][]{1998A&A...330...79L,2005ApJ...619...73H}. Using a
``time-lag core shift'' method, which proves to be a good alternative to VLBI 
core-shifts \citep[e.g.][]{2011MNRAS.415.1631K, 2014MNRAS.437.3396K}, the 
positions of the nine unit-opacity surfaces (or photospheres) with 
respect to the jet base were inferred. 
These are:
$(10.2 \pm 1.2)$\,pc for the 2.64\,GHz core,  
$ (7.1 \pm 1.0)$\,pc at 4.85,
$ (5.3 \pm 0.8)$\,pc at 8.35, 
$ (4.8 \pm 0.8)$\,pc at 10.45,
$ (4.0 \pm 0.7)$\,pc at 14.60,  
$ (2.6 \pm 0.6)$\,pc at 23.05,  
$ (2.9 \pm 1.0)$\,pc at 32.00,  
$ (3.4 \pm 0.8)$\,pc at 43.05, and  
$ (4.0 \pm 1.1)$\,pc at 86.24\,GHz (see Table 4 of Ref. 
\cite{2016A&A...590A..48K} for the values and their calculation).
Finally, we obtain the equipartition magnetic field along the jet axis, with 
values increasing from 14\,mG at the 2.64\,GHz core up to 176\,mG at 86.24\,GHz.

\subsection{Where do the $\gamma$ rays come from?}

The location of the $\gamma$-ray emitting region is decisively constrained by 
combining the aforementioned findings with those of \cite{2014MNRAS.441.1899F} 
and \cite{2016A&A...586A..60K}. Specifically, in \cite{2014MNRAS.441.1899F} the 
authors calculate the relative distance between the 86\,GHz core and the 
$\gamma$-ray site to be about 2.1\,pc, while in \cite{2016A&A...586A..60K}, 
using high-resolution mm-VLBI imaging, the knot most likely 
connected with the flare in question is identified (knot C3) and its 
kinematical properties are deduced ($\beta_{\rm app} \sim 7$\,c, jet viewing 
angle of $2.6^{\circ}$).

Having calculated the absolute distance between the jet base and the 86.24\,GHz 
core $(4.0 \pm 1.1)$\,pc, and with the relative separation between the same 
core and the $\gamma$-ray site (2.1\,pc), the high-energy emission  
originates at  $(1.9 \pm 1.1)$\,pc from the jet base. This places it far from 
the bulk of the broad-line material of the source (${\rm R}_{\rm BLR} \approx 
0.1$\,pc) making external Compton, on a field other than that of the 
BLR, a very relevant emission mechanism for 
PKS\,1502+106's $\gamma$ ray emission, with implications on the origin of the 
seed photon field (see \cite{2016A&A...590A..48K} and references therein).

%%%%%%%%%%%%%%%%%%%%%%%%%%%%%%%%%%%%%%%%%%
% \section{Conclusions}
% 
% This section is not mandatory, but can be added to the manuscript if the 
% discussion is unusually long or complex.

%%%%%%%%%%%%%%%%%%%%%%%%%%%%%%%%%%%%%%%%%%
% \vspace{6pt} 

%%%%%%%%%%%%%%%%%%%%%%%%%%%%%%%%%%%%%%%%%%
% %% optional
% \supplementary{The following are available online  at www.mdpi.com/link, 
% Figure S1: title, Table S1: title, Video S1: title.}

%%%%%%%%%%%%%%%%%%%%%%%%%%%%%%%%%%%%%%%%%%
\acknowledgments{V.K. thanks L. Fuhrmann, E. Angelakis, T. P. Krichbaum, I. 
Myserlis, I. Nestoras,  J. A. Zensus, and H. Ungerechts for interesting 
discussions and useful comments; also W. Max-Moerbeck and B. Boccardi for 
carefully reading the paper. Partially based on observations with the 
Effelsberg 100-m 
% telescope 
% of the MPIfR 
% at  
and the IRAM 30-m telescopes.
% at Pico Veleta.
}

\def\memsai{Memorie della Società Astronomica Italiana}
\def\aj{AJ}%
          % Astronomical Journal
\def\actaa{Acta Astron.}%
          % Acta Astronomica
\def\araa{ARA\&A}%
          % Annual Review of Astron and Astrophys
\def\apj{ApJ}%
          % Astrophysical Journal
\def\apjl{ApJ}%
          % Astrophysical Journal, Letters
\def\apjs{ApJS}%
          % Astrophysical Journal, Supplement
\def\ao{Appl.~Opt.}%
          % Applied Optics
\def\apss{Ap\&SS}%
          % Astrophysics and Space Science
\def\aap{A\&A}%
          % Astronomy and Astrophysics
\def\aapr{A\&A~Rev.}%
          % Astronomy and Astrophysics Reviews
\def\aaps{A\&AS}%
          % Astronomy and Astrophysics, Supplement
\def\azh{AZh}%
          % Astronomicheskii Zhurnal
\def\baas{BAAS}%
          % Bulletin of the AAS
\def\bac{Bull. astr. Inst. Czechosl.}%
          % Bulletin of the Astronomical Institutes of Czechoslovakia
\def\caa{Chinese Astron. Astrophys.}%
          % Chinese Astronomy and Astrophysics
\def\cjaa{Chinese J. Astron. Astrophys.}%
          % Chinese Journal of Astronomy and Astrophysics
\def\icarus{Icarus}%
          % Icarus
\def\jcap{J. Cosmology Astropart. Phys.}%
          % Journal of Cosmology and Astroparticle Physics
\def\jrasc{JRASC}%
          % Journal of the RAS of Canada
\def\mnras{MNRAS}%
          % Monthly Notices of the RAS
\def\memras{MmRAS}%
          % Memoirs of the RAS
\def\na{New A}%
          % New Astronomy
\def\nar{New A Rev.}%
          % New Astronomy Review
\def\pasa{PASA}%
          % Publications of the Astron. Soc. of Australia
\def\pra{Phys.~Rev.~A}%
          % Physical Review A: General Physics
\def\prb{Phys.~Rev.~B}%
          % Physical Review B: Solid State
\def\prc{Phys.~Rev.~C}%
          % Physical Review C
\def\prd{Phys.~Rev.~D}%
          % Physical Review D
\def\pre{Phys.~Rev.~E}%
          % Physical Review E
\def\prl{Phys.~Rev.~Lett.}%
          % Physical Review Letters
\def\pasp{PASP}%
          % Publications of the ASP
\def\pasj{PASJ}%
          % Publications of the ASJ
\def\qjras{QJRAS}%
          % Quarterly Journal of the RAS
\def\rmxaa{Rev. Mexicana Astron. Astrofis.}%
          % Revista Mexicana de Astronomia y Astrofisica
\def\skytel{S\&T}%
          % Sky and Telescope
\def\solphys{Sol.~Phys.}%
          % Solar Physics
\def\sovast{Soviet~Ast.}%
          % Soviet Astronomy
\def\ssr{Space~Sci.~Rev.}%
          % Space Science Reviews
\def\zap{ZAp}%
          % Zeitschrift fuer Astrophysik
\def\nat{Nature}%
          % Nature
\def\iaucirc{IAU~Circ.}%
          % IAU Cirulars
\def\aplett{Astrophys.~Lett.}%
          % Astrophysics Letters
\def\apspr{Astrophys.~Space~Phys.~Res.}%
          % Astrophysics Space Physics Research
\def\bain{Bull.~Astron.~Inst.~Netherlands}%
          % Bulletin Astronomical Institute of the Netherlands
\def\fcp{Fund.~Cosmic~Phys.}%
          % Fundamental Cosmic Physics
\def\gca{Geochim.~Cosmochim.~Acta}%
          % Geochimica Cosmochimica Acta
\def\grl{Geophys.~Res.~Lett.}%
          % Geophysics Research Letters
\def\jcp{J.~Chem.~Phys.}%
          % Journal of Chemical Physics
\def\jgr{J.~Geophys.~Res.}%
          % Journal of Geophysics Research
\def\jqsrt{J.~Quant.~Spec.~Radiat.~Transf.}%
          % Journal of Quantitiative Spectroscopy and Radiative Trasfer
\def\memsai{Mem.~Soc.~Astron.~Italiana}%
          % Mem. Societa Astronomica Italiana
\def\nphysa{Nucl.~Phys.~A}%
          % Nuclear Physics A
\def\physrep{Phys.~Rep.}%
          % Physics Reports
\def\physscr{Phys.~Scr}%
          % Physica Scripta
\def\planss{Planet.~Space~Sci.}%
          % Planetary Space Science
\def\procspie{Proc.~SPIE}%
          % Proceedings of the SPIE
%
\bibliographystyle{mdpi}
\bibliography{references.bib}
%%%%%%%%%%%%%%%%%%%%%%%%%%%%%%%%%%%%%%%%%%
%% optional
% \sampleavailability{Samples of the compounds ...... are available from the authors.}
%%%%%%%%%%%%%%%%%%%%%%%%%%%%%%%%%%%%%%%%%%
\nopagebreak
\end{document}